# On the interpretation of the He II absorption in the line of sight of Q0302-003

Biman B. Nath and Shiv K. Sethi
*Inter-University Centre for Astronomy and Astrophysics, Post Box 4, Ganeshkhind, Pune – 411007, India*

**ABSTRACT**
We point out the peculiarities of the line of sight of Q0302-003 that was recently used to estimate the He II opacity at $z \sim 3.285$. This line of sight contains a large void in Lyman $\alpha$ clouds, in the region used for the He II opacity observation, and the void is thought to be caused by the ionizing radiation from a foreground quasar (Dobrzycki and Bechtold 1991). If this is true, then the radiation from the foreground quasar is more relevant in the region used for the estimation of the He II opacity than the diffuse UV background radiation. We argue that in that case the observed opacity should not be used to put constraint on the diffuse UV background radiation. Using a typical spectral index of quasars, we estimate that most of the observed He II opacity must be due to Gunn-Peterson effect and suggest a H I Gunn-Peterson test along this line of sight. We further discuss the clumpiness of the intergalactic medium in the vicinity of Q0302-003 as suggested by observations.

## 1 INTRODUCTION

One of the most important endeavours in observational cosmology has been the attempt to detect the diffuse component of the intergalactic medium (IGM). In the standard cosmological model, according to which the universe recombined at redshifts $\sim 1100$, one expects to see absorption troughs in the spectra of high redshift quasars due to neutral atoms in their lines of sight (Gunn and Peterson 1965). Various upper limits on, and a few recent, tentative detections of this 'Gunn-Peterson' (GP) absorption due to hydrogen atoms put severe constraints on the H I content of the diffuse IGM. In the model of a photoionized universe, consequently, these observations constrain the intensity and spectral shape of the UV background radiation at high redshift. The recent detection of Lyman $\alpha$ absorption due to He II atoms ($\tau_{Ly\alpha}^{\text{He II}}$) in the line of sight of the quasar Q0302-003 by Jakobsen et al.(1994) has, therefore, stirred discussions on the nature of the background ionizing radiation in the high redshift universe — since the spectrum of the radiation should be soft enough to ionize most of the H I but not much of the He II atoms (Madau and Meiksin 1994).

However, we found that the above mentioned quasar Q0302-003 at $z_{em} \sim 3.285$ (Q1, hereafter) was previously observed, among others, by Dobrzycki and Bechtold (1991), and was noted by them for a large ($\sim 10$ Mpc) void in the Lyman $\alpha$ clouds in its line of sight. They noticed that the void was close to another nearby quasar in the foreground, Q0301-005 at $z_{em} \sim 3.223$ (Q2, hereafter), and discussed the possibility that the void was caused by the ionizing radiation from this second quasar, Q2. This void accounts for about 30% of the redshift space used by Jakobsen et al.(1994) to derive the value of $\tau_{Ly\alpha}^{\text{He II}}$ at $z \sim 3.3$. Now, if the quasar Q2 is indeed responsible for this void, the effect of the ionizing radiation from this quasar must be more important in this region than the global ionizing radiation field (denoted by $J_\nu$). Consequently, the interpretation of the observed optical depth must be more relevant for the spectrum of the local ionizing radiation than that of $J_\nu$.

In this paper, we discuss the importance of taking the ionizing radiation from Q2 into account in interpreting the $\tau_{Ly\alpha}^{\text{He II}}$ detected by Jakobsen et al.(1994). Instead of using this opacity to derive constraints on $J_\nu$, we use the spectral shape of the local ionizing radiation to estimate the maximum contribution to the He II optical depth due to line blanketing. We then discuss upon the physical state of the diffuse IGM in the vicinity of the quasars Q1 and Q2.

## 2 THE VOID AND THE REGION NEAR Q1 AND Q2

Jakobsen et al.(1994) used the flux decrement in a 50Å bin shortward of the He II Lyman $\alpha$ line ($\lambda_{rest} = 304$Å) in their data to derive a value of $\tau_{Ly\alpha}^{\text{He II}} \approx 3.2_{+\infty}^{-1.1}$, and a 90% lower bound of $\sim 1.7$ on $\tau_{Ly\alpha}^{\text{He II}}$. For the 304Å line, and for $z_{em}$ of the quasar Q1 equal to 3.285, this bin corresponds to a redshift space between $z_{min} \sim 3.12$ and $z_{em}$. In a $q_o = 0.5$ universe, this means a physical distance $\sim 27 h_{50}^{-1}$ Mpc in the line of sight of Q1 (where the Hubble constant is $H_o = 50 h_{50}$ km s$^{-1}$ Mpc$^{-1}$; we will use $q_o = 0.5$ throughout the paper).

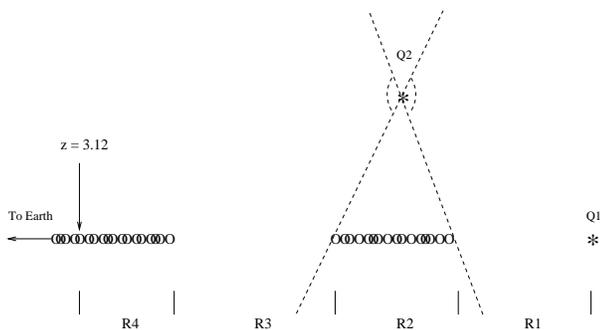

**Figure 1.** Schematic diagram of the region in the vicinity of Q0302-003 (Q1) and Q0301-005 (Q2), drawn to scale for $q_o = 0.5$ (see also Fig. 3 of DB91). The arrow at $z = 3.12$ marks the data bin ($z = 3.285 - 3.12$) used to estimate the He II opacity by Jakobsen et al.(1994). The distribution and the sizes of the Lyman $\alpha$ clouds are only schematic. The void referred to in the text occupies the region R3 marked in here.

Dobrzycki and Bechtold (1991) (hereafter referred to as DB91), observing the same quasar, detected a void in the Lyman $\alpha$ lines, between redshifts $3.15 \lesssim z \lesssim 3.20$ in its line of sight, corresponding to a proper size of $\sim 8.3 h_{50}^{-1}$ Mpc for the void. They estimated the probability of seeing such a void by chance to be $\sim 2 \times 10^{-4}$. They discussed another possible void nearer to Q1, claiming that the two lines (see the spectrum in their Fig. 1) in that region could be metal lines of an absorption system at low redshift. However, Bechold (1994) listed these two lines as Lyman $\alpha$ lines, and we will discuss this particular void as being only tentative.

DB91 argued that the void at around $z \sim 3.17$ could have been caused by the ionizing radiation from a nearby quasar Q2 in foreground ($z_{em} = 3.223$), which is separated from the line of sight of Q1 by $17'$ (7.5 Mpc) (see their Fig. 3). Since Q2 is not centered on the void, one possibility is that the radiation from Q2 is limited to radiation cones, with a beaming angle of $\sim 140^\circ$. DB91 further calculated the luminosity $F_\nu$ of Q2 at the centre of the void at the Lyman limit. This essentially puts a limit on the intensity of the diffuse ionizing background at the Lyman limit, and this was derived as $J_\nu \lesssim (\frac{F_\nu}{4\pi}) \sim 10^{-21.8}$ erg s$^{-1}$ cm$^{-2}$ Hz$^{-1}$ sr$^{-1}$ (independent of $H_o$). Since this value of $J_{912}$ at the hydrogen Lyman limit is rather low compared to $10^{-21}$ erg s$^{-1}$ cm$^{-2}$ Hz$^{-1}$ sr$^{-1}$ estimated in other lines of sight at $z \sim 3$, they argued that this could be either due to the fact that Q2 radiates more in the direction of the void for some reason, or there are some other extra source(s) of ionizing radiation, or the local value of $J_{912}$ is indeed small in this region (see also Bechtold 1994).

In Figure 1, we draw a schematic diagram of this region, following Fig. 3 of DB91, and mark the above mentioned redshift space used in the observation of He II GP effect by Jakobsen et al.(1994) ($z = 3.285 - 3.12$). It is seen that the void accounts for about 30% of the data bin used to estimate the $\tau_{Ly\alpha}^{\text{He II}}$ (marked as region R3 in Fig. 1), and, subsequently, to put constraints on the ratio of $J_{912}$ to $J_{228}$ (at hydrogen and helium Lyman limit) (Madau and Meiksin 1994; hereafter, MM94). However, if the void is taken into account, whose very existence implies the importance of the ionizing radiation from Q2 over the diffuse background $J_\nu$, at least in the region of the void, then it does not seem reasonable to use Jakobsen et al.'s result to put constraints on the spectrum of $J_\nu$. In physical space, the midpoint of Jakobsen et al.'s data bin of $50 \text{\AA}$ is close to the edge of the void nearer to Q1. If the radiation cone hypothesis of Dobrzycki and Bechtold (1991) (shown in Fig. 1) is correct, then the intrinsic ionizing radiation from Q2 and Q1 is also important in region R1, the second, tentative void detected by DB91 (which amounts to $\sim 25\%$ of Jakobsen etal 's data bin).

## 3 INTERPRETING THE HE II OPACITY

### 3.1 Line blanketing and He II Gunn-Peterson test

MM94 estimated the He II opacity due to line blanketing from (He II atoms in) Lyman $\alpha$ clouds, using a average distribution of H I equivalent widths of Lyman $\alpha$ clouds, and calculating the mean He II equivalent width from a curve of growth analysis, for a given value of $S_L$. For a gas in photoionization equilibrium with an ionizing background radiation with 'softness' parameter $S_L$, the column densities of He II and H I are related as $N_{\text{He II}} \sim 1.8 S_L N_{\text{H I}}$. They argued that line blanketing itself could account for the lower bound of the He II opacity ($\sim 1.7$), if $S_L > 200$, or $S_L > 40$, depending on whether the lines are broadened by thermal or bulk velocities, respectively.

Furthermore, they used an upper bound on $\tau_{GP}^{\text{H I}}$ at $z \sim 3$ (in the line of sight of a different quasar), and the fact that $\tau_{GP}^{\text{He II}}$ and $\tau_{GP}^{\text{H I}}$ are related by a factor proportional to $S_L$ (their eqn. 3), to estimate $\tau_{GP}^{\text{He II}}$. Then they put constraints on $S_L$ in the case that both line blanketing and $\tau_{GP}^{\text{He II}}$ contribute towards the total He II opacity. For thermally broadened lines, e.g., they arrived at a lower bound of $S_L > 30$.

In view of our discussion in the previous section, on the peculiarity of the local region near Q1, i.e., the presence of the void and that of Q2 nearby, the ionizing radiation from Q2 is more important than the global radiation field $J_\nu$, in regions covering $\sim 30\%$ of the data bin used for He II opacity observation. A detail analysis should divide the region into four parts, as in Fig. 1, and discuss the implication of the He II opacity in each of them, taking into account the sources of the ionizing radiation that is appropriate for each region. Since Jakobsen et al.'s data had an accuracy of $\sim 10 - 20 \text{\AA}$, which is $\sim 20 - 40\%$ of the whole bin, it is however not possible to analyse and interpret the data in such detail. As the goal of this paper is to point out the importance of taking the local radiation field into account, we instead chose to demonstrate this by taking the observed spectral shape of Q2, and its $S_L$, which is at least important in the void, if not anywhere else along the line of sight, and which covers a substantial fraction of the data bin used by Jakobsen et al.(1994).

We emphasize that our analysis will not give the exact picture. The resolution of Jakobsen et al.'s data does not permit one do such an exercise. However, our goal is to draw attention to the peculiarity of the line of sight used for He II opacity observation, and to argue that one needs to be cautious before reaching conclusions about the global radiation field of $J_\nu$, we will here show the differing conclusions one reaches after using the local radiation field as the known parameter.

Sargent, Steidel and Boksenberg (1989) determined the spectral index of Q2 as being $\alpha \sim 0.89$ ($f_\nu \propto \nu^{-\alpha}$), in the visible part of the spectrum. However, the success rate of finding $z > 3$ quasars in UV indicate that the typical 'visible to EUV' spectral index of quasars is $\sim 1.5$, corresponding to $S_L \sim 8$, (P. Jakobsen, private communication). We will use this value of $S_L$ for Q2 in the following.

Here, we note that the spectrum of the ionizing radiation from Q2 as observed in the void could be different from the intrinsic radiation from Q2 because of absorption by matter between Q2 and the void. Q2 is separated from the centre of the void by $\sim 10.5 h_{50}^{-1}$ Mpc, and from the midpoint of the physical space between $z = 3.12$ and $z = 3.285$, by $\sim 11 h_{50}^{-1}$ Mpc. From the observed redshift distribution of Lyman limit systems (Lanzetta 1991), we estimate a probability of $\sim 0.1$ for a Lyman limit system to be there between Q2 and either of these points. H I, He III or He II atoms in the diffuse gas in the region can also change $S_L$ of Q2's spectrum before the radiation reaches the void. Observations at $z \sim 3$ however suggests the absence of any substantial amount of H I atoms, and it is unlikely in a photoionized IGM that the abundance of He III atoms would be substantial. For He II atoms, if the result of Jakobsen et al.(1994) is taken at its face value, i.e., $\tau_{GP}^{\mathrm{He\,II}} \gtrsim 1.7$, a lower limit of $y = \frac{n_{\mathrm{He\,II}}}{n_{\mathrm{He}}} \gtrsim 10^{-3} h_{50}^{-1} (\frac{\Omega_{\mathrm{IGM}}}{0.05})^{-1}$ is implied. The optical depth for a photon with $\lambda = 228\mathring{A}$ for the distance between Q2 and the void is $\tau \sim 0.3 (\frac{y}{10^{-3}})(\frac{\Omega_{\mathrm{IGM}}}{0.05}) h_{50}^2$. In view of such a small optical depth, it is reasonable to assume that the ratio of intensities of radiation from Q2, at $912\mathring{A}$ and $228\mathring{A}$, is largely unchanged from the intrinsic one. The value of $y$ is unlikely to be much greater than this, even in the case where the known quasars are the only sources of $J_\nu$, as shown by Madau (1992) ($y \propto \tau_{GP}^{\mathrm{He\,II}} \propto J_\nu^{-1}$). If it is much larger, then our analysis here will not be valid. However, it must be admitted that absorption by He II could increase the value of $S_L$.

The He II optical depth due to line blanketing can then be calculated for $S_L \sim 8$, by using the curve of growth, and using the expression for effective optical depth averaged over all lines of sight (Paresce, McKee and Bowyer 1980),

$$\tau_{eff}(z) = \frac{1+z}{\lambda_\alpha} \int_{W_{min}}^{W_{max}} \frac{\partial^2 N}{\partial W \partial z} W \, dW \qquad (1)$$

where $\lambda_\alpha$ is the rest wavelength, and $\partial^2 N/\partial W \partial z$ is the rest wavelength distribution of the absorbers. We use the same distribution of H I rest wavelengths as in MM94, i.e., one that was derived by Murdoch et al.(1986), as

$$\begin{aligned}\frac{\partial^2 N}{\partial W \partial z} &= 40.7 \exp\left(-\frac{W^{\mathrm{H\,I}}}{W_*}\right)(1+z)^{2.46} \\ & \qquad (0.2 < W^{\mathrm{H\,I}} < 2 \mathring{A}); \\ &= 11.4 \left(\frac{W^{\mathrm{H\,I}}}{W_*}\right)^{-1.5}(1+z)^{2.46} \\ & \qquad (W_{min}^{\mathrm{H\,I}} < W^{\mathrm{H\,I}} < 0.2 \mathring{A}), \qquad (2)\end{aligned}$$

where $W_* = 0.3\mathring{A}$. We use a Doppler parameter of $b^{\mathrm{H\,I}} = 35$ km s$^{-1}$ for all the clouds (as in MM94). We derive the line blanketing opacity in He II as $\tau_{eff}^{\mathrm{He\,II}}(S_L = 8) \sim 0.57 (W_{min}^{\mathrm{H\,I}} = 0.01 \mathring{A})$, for velocity broadened clouds (i.e., $b^{\mathrm{He\,II}} = b^{\mathrm{H\,I}}$). For thermally broadened clouds ($b^{\mathrm{He\,II}} = 0.5 b^{\mathrm{H\,I}}$) we estimated $\tau_{eff}^{\mathrm{He\,II}}(S_L = 8) \sim 0.41$, for the same lower limit in $W^{\mathrm{H\,I}}$. Decreasing the lower limit even to $W_{min}^{\mathrm{H\,I}} = 0.0025$ (corresponding to $N_{\mathrm{H\,I}} \sim 10^{12}$ cm$^{-2}$; see Songaila, Hu and Cowie 1995) does not change the opacities much for such low values of $S_L$ (see Fig. 1 of MM94). These numbers are also consistent with the curves in Jakobsen et al.(1994). Compared to the 90% lower bound on $\tau_{Ly\alpha}^{\mathrm{He\,II}}$ of $\sim 1.7$, the line blanketing opacity for $S_L = 8$ is, therefore, small. Moreover, due to the presence of the void in Lyman $\alpha$ clouds, the line blanketing opacity is expected to be even smaller than the above estimates.

Therefore, if the radiation near the void is representative of that in the whole region used by Jakobsen et al.(1994), then most of the observed He II opacity must come from He II Gunn-Peterson effect. In any case, because of the presence of the void, one can certainly say that the constraints on the $S_L$ for the diffuse background radiation, $J_\nu$, are much lower than those in MM94.

### 3.2 H I Gunn-Peterson Test

If most of the observed He II opacity is due to $\tau_{GP}^{\mathrm{He\,II}}$, then one can readily estimate the Gunn-Peterson opacity due to hydrogen in the IGM, $\tau_{GP}^{\mathrm{H\,I}}$, along this line of sight. The two opacities are related as $\tau_{GP}^{\mathrm{H\,I}} = 4\tau_{GP}^{\mathrm{He\,II}}(\frac{N_{\mathrm{H\,I}}}{N_{\mathrm{He\,II}}}) = 4\tau_{GP}^{\mathrm{He\,II}}(\frac{1}{1.8 S_L})$. For $S_L = 8$, $\tau_{GP}^{\mathrm{He\,II}} \sim 1.1$, and $\tau_{GP}^{\mathrm{H\,I}} \sim 0.31$, if line blanketing from velocity broadened clouds is considered. If clouds are thermally broadened clouds, $\tau_{GP}^{\mathrm{H\,I}} \sim 0.36$. However, $S_L$ of Q2 at the void could be somewhat more than its intrinsic value, as we discussed in section 3.1. For, say, $S_L = 15$, $\tau_{GP}^{\mathrm{H\,I}} \sim 0.13$ (0.17 for thermally broadened clouds). Such a value of $\tau_{GP}^{\mathrm{H\,I}}$ at $z \sim 3.3$ is consistent with, say, the limits ($\tau_{GP}^{\mathrm{H\,I}} \lesssim 0.115 \pm 0.025$, at $z \sim 3.4$) put by Fang and Crotts (1995) (using a method which is claimed to be better than the previous ones). We note that such a value of $S_L$ is below the lower limits put by MM94.

Turning the argument around, we can say that the determination of $\tau_{GP}^{\mathrm{H\,I}}$ along this line of sight would give the value of $S_L$ (or, at least a strict lower limit on $S_L$, from an upper limit on $\tau_{GP}^{\mathrm{H\,I}}$). This will readily test the ideas presented here. If $S_L$ is indeed very large (MM94, Songaila, Hu and Cowie 1995), then $\tau_{GP}^{\mathrm{He\,II}}$ will be very small, and consequently, $\tau_{GP}^{\mathrm{H\,I}}$ will also be very small along this line of sight. The recent observations for H I Gunn-Peterson test have, however, yielded disparate results (see, e.g., Reisenegger and Miralda-Escudè 1995). Therefore, we suggest a H I Gunn-Peterson test along this particular line of sight instead of using the estimates for $\tau_{GP}^{\mathrm{H\,I}}$ from other lines of sight.

### 3.3 Clumpiness of intergalactic medium

If one takes note of the rather low value of $J_{912}$ in the line of sight of Q1, as estimated by DB91, then one can speculate upon the density of the IGM in the vicinity of Q1 and Q2. We bear in mind here the caveat that various effects, including special beaming configuaration of Q2 and the presence of other unknown sources nearby, could increase this estimate of $J_{912}$. The $\tau_{GP}^{\rm He\,II}$, for a (diffuse) gas density of the IGM equal to $\Omega_{\rm IGM}$ at redshift $z$, is given by (see, *e.g.*, MM94)

$$\tau_{GP}^{\rm He\,II}(z) \sim 25 \left(\frac{1+z}{4.3}\right)^{4.5} f \Omega_{\rm IGM}^2 h_{50}^3 S_L J_{912,-21}^{-1}, \quad (3)$$

where $J_{912} = J_{912,-21}\, 10^{-21}$ erg cm$^{-2}$ s$^{-1}$ sr$^{-1}$ Hz$^{-1}$. We have included a fudge factor $f$ for the local IGM here, which is appropriate if $\Omega_{\rm IGM}$ is not a constant throughout the universe, and can be thought in terms of the clumpiness of the IGM, with $f = \frac{\langle \rho^2 \rangle}{\langle \rho \rangle^2}$ (Miralda-Escudè and Ostriker 1990). Using $S_L \sim 3.43$, $\tau_{GP}^{\rm He\,II} \sim 1$, and $J_{912,-21} \lesssim 0.16$ (for $q_o = 0.5$, DB91), the above expression for $\tau_{GP}^{\rm He\,II}$ can then be inverted to obtain

$$f \Omega_{\rm IGM}^2 h_{50}^3 \quad \gtrsim \quad 8 \times 10^{-4} \left(\frac{J_{912,-21}}{0.16}\right)\left(\frac{S_L}{8}\right)^{-1}$$
$$\left(\frac{\tau_{GP}^{\rm He\,II}(z=3.3)}{1}\right). \quad (4)$$

The estimated baryon content of the universe from nucleosynthesis is $\Omega_b^2 h_{50}^4 \sim 2.5 \times 10^{-3}$ (Walker et al.1991). Thus, for $h_{50} = 1$, a value of $f \gtrsim 1$ is indicated, for $\Omega_{\rm IGM} < \Omega_b$.

The observed He II opacity along the line of sight of Q1, therefore, may suggest a locally overdense region, if the above estimates of $S_L$, $J_{912,-21}$, and $\tau_{GP}^{\rm He\,II}$ are correct and representative of the redshift space used by Jakobsen et al.(1994) in their observation. (Note that the void in the Lyman $\alpha$ clouds does not mean an underdense intergalactic medium; it is simply due to the proximity effect from the ionizing radiation from Q2.)

### 4 SUMMARY AND DISCUSSIONS

If the line of sight towards Q1 is not a representative one, as we have argued, then observations of He II opacity along other lines of sight, even at the same redshift, is bound to yield different values of opacity and $\tau_{GP}^{\rm He\,II}$. Recently, Reisenegger and Miralda-Escudè (1995) have discussed the widely differing results from H I Gunn-Peterson tests conducted at same redshifts, and ascribed the differences to the fact that $\Omega_{\rm IGM}$ is not homogeneous. They argued that, taken separately, some of the observed estimates of $\tau_{GP}^{\rm H\,I}$ can lead one to conclusions about the global IGM that may not necessarily be correct.

It is interesting to note that a few other close pairs of quasars have been recently observed, and the value of $J_\nu$, estimated from the voids near the foreground quasars (Fernández-Soto et al.1995). The ideas presented in this work could in principle be tested by comparing the $\tau_{Ly\alpha}^{\rm He\,II}$ along such lines of sight and that in a line of sight without any foreground quasar. Unfortunately, however, the quasars observed by them have redshifts too low ($z < 2.7$) for the He II opacity observations using the Hubble Space Telescope.

Here we have pointed out the importance of being cautious in putting constraints on the global UV background radiation, or the global IGM, from what is essentially a piece of observation along just one line of sight, and that too, a rather peculiar one, with voids in Lyman $\alpha$ clouds and a foreground quasar Q2. We have instead used a typical quasar UV spectrum for Q2, and assuming that it is typical in the region near Q1 and Q2, we have shown that the bounds on the diffuse UV background radiation is smaller than that has been claimed by previous workers. The complicated configuration of this region, combined with the low resolution of the He II opacity observations, makes it difficult to work out the precise constraints. We have, therefore, only attempted to point out the peculiarity of this particular line of sight, and the fact that any discussion on the observed He II opacity will be incomplete without taking the local radiation field into account.

We showed that this exercise suggests that (a) most of the contribution to the He II opacity of Jakobsen et al.(1994) comes from $\tau_{GP}^{\rm He\,II}$; (b) the existence of the void in Lyman $\alpha$ clouds in the line of sight shows the importance of the ionizing radiation from the foreground quasar along this line of sight, and argues against the use of the observed He II opacity to put constraints on the global diffuse UV background radiation in the universe; (c) the IGM in the vicinity of the two quasars is probably overdense compared to the global average of $\Omega_{\rm IGM}$. We have suggested H I Gunn-Peterson test in this line of sight to test the ideas presented here.

### Acknowledgements

We are indebted to Dr. Peter Jakobsen and an anonymous for their valuable comments on the manuscript. We also thank Drs. Jill Bechtold, Puragra Guhathakurta for their suggestions.